\documentclass[12pt]{article}
\usepackage{amsmath,bm,amsfonts,amssymb}

\renewcommand{\baselinestretch}{1.1}
\newcommand{\CP}[1]{\mathbb{P}^{#1}}
\newcommand{\C}[1]{\mathbb{C}^{#1}}

\newcommand{\R}[1]{\mathbb{R}^{#1}}

\newcommand{\WP}[1]{{\mathbb{P}({#1})}}
\def\lcm{\mathrm{lcm}}

\def\Z{\mathbb{Z}}

\def\Y{\mathbf{Y}}
\def\S{\mathbf{S}}
\def\T11{{T}^{1,1}}
\def\be{\begin{equation}}
\def\ee{\end{equation}}
\def\bear{\begin{eqnarray}}
\def\eear{\end{eqnarray}}

\def\bz{\mathbf{z}}
\def\bV{\mathbf{V}}

\def\bX{\mathbf{X}}

\def\Vol{\mathrm{Vol}}
\def\vol{\mathrm{vol}}
\def\bg{\mathbf{g}}
\def\bh{\mathbf{h}}

\def\bR{\mathbf{R}}

\def\Tr{\mathrm{Tr}}

\def\bi{\bibitem}
\def\map#1.#2.{#1 \longrightarrow #2}
\def\pr #1.{\mathbb P^{#1}}
\def\Hz #1.{\mathbb F_{#1}}
\def\llist#1.#2.{{#1}_1,{#1}_2,\dots,{#1}_{#2}}
\usepackage{graphicx}

\begin{document}

\begin{titlepage}

\begin{flushright}
hep-th/0305048\\
NSF-KITP-03-27
\end{flushright}
\vfil

\begin{center}
{\huge Dibaryon Spectroscopy }\\
\vspace{3mm}
%{\huge }

\end{center}

\vfil
\begin{center}
{\large Christopher P. Herzog$^\sharp$ and 
James M\raise 4.5pt \hbox{\text {\normalsize c}}Kernan$^\flat$}\\
\vspace{1mm}
$\sharp$ KITP, UCSB, \; \; 
$\flat$ Dept. of Math., UCSB, \\
Santa Barbara, CA  93106, U.S.A.\\
{\tt herzog@kitp.ucsb.edu,
mckernan@math.ucsb.edu}\\
\vspace{3mm}
\end{center}

\vfil

\begin{center}
{\large Abstract}
\end{center}

\noindent
The AdS/CFT correspondence relates
dibaryons
in 
%${\mathcal N}=1$ 
superconformal gauge theories to 
holomorphic curves in K\"{a}hler-Einstein surfaces.  
The degree of the holomorphic curves
is proportional to the gauge theory conformal dimension
of the dibaryons.  Moreover, the number of 
holomorphic curves should match,
in an appropriately defined sense, the number of dibaryons.
%We calculate the volumes of three-cycles in
%a large class of five dimensional Einstein-Sasaki
%manifolds.  These volumes are important in the
%context of the AdS/CFT correspondence because D3-branes
%wrapping these cycles are conjectured to correspond
%to dibaryonic operators in corresponding quiver
%gauge theories.  
Using AdS/CFT backgrounds built from the generalized
conifolds of Gubser, Shatashvili, and Nekrasov (1999),
we show that the gauge theory
prediction for the dimension of dibaryonic operators 
does indeed match the degree of the corresponding holomorphic
curves.
For AdS/CFT backgrounds built from cones over del Pezzo 
surfaces,
we are able to match the degree of the curves to the conformal
dimension of dibaryons for the
$n$th del Pezzo surface, $1\leq n \leq 6$.
Also, for the del Pezzos and the $A_k$ type generalized conifolds,
for the dibaryons of smallest conformal dimension, we are 
able to match the number of holomorphic curves with the number of
possible dibaryon operators from gauge theory.
\vfil
\begin{flushleft}
May 2003
\end{flushleft}
\vfil
\end{titlepage}
\newpage
\renewcommand{\baselinestretch}{1.1}  %looks better

%%%%%%%%%%%%%%%%%%%%%%%%%%%%%%%%%%%%%%%%%%%%%
%% include the next line for double spacing %%
%%%%%%%%%%%%%%%%%%%%%%%%%%%%%%%%%%%%%%%%%%%%%%
%\renewcommand{\baselinestretch}{2}

\section{Introduction}

AdS/CFT correspondence 
\cite{jthroat, gkp, EW}
asserts that type IIB string theory
on $AdS_5 \times \bX$ is equivalent to a 
superconformal
quiver gauge theory.  The geometric objects involved here are
five dimensional anti-de Sitter space $AdS_5$ and a
five dimensional Sasaki-Einstein manifold $\bX$.  
A supersymmetric quiver gauge theory contains a collection
of vector multiplets transforming under $SU(N)$ gauge groups
for each node in the quiver 
and, for each line in the quiver, a collection of chiral multiplets 
transforming
under bifundamental and adjoint
representations of the gauge groups.
Despite extensive efforts,
AdS/CFT remains a conjecture and the present work was motivated
largely by the need to find checks of the correspondence which
do not require explicit knowledge of the metric on $\bX$.

To review the origins of AdS/CFT, recall that the original 
correspondence
\cite{jthroat, gkp, EW} was motivated by comparing a stack
of $N$ elementary branes with the metric it produces
(for reviews, see for example \cite{magoo, Krev}).  
For a stack placed in ten dimensional space, the theory
on the branes is ${\mathcal N}=4$ supersymmetric
$SU(N)$ gauge theory.  The quiver would have a single node,
corresponding to a single ${\mathcal N}=4$ vector
multiplet.
The gravitational back reaction from the
stack
causes the space to factorize into $AdS_5 \times \S^5$
close to the D3-branes; the gauge theory on 
the branes is conjectured to be equivalently described
by type
IIB string theory in this factorized background.

In order to break some of the supersymmetry (SUSY),
we may place
the branes at a conical singularity instead of in flat ten
dimensional space \cite{LNV, KS, KW, Kehag, MoPle, Acharya}.
For example, branes placed at the orbifold
singularity of $\C{2}/\Gamma$ where $\Gamma$ is 
a discrete subgroup of $SU(2)$ preserve ${\mathcal N}=2$
SUSY \cite{LNV, KS}.
The gauge theory quivers correspond to
simply laced Dynkin diagrams, as will be 
reviewed in greater detail in section 5.  
Geometrically, the ten dimensional space
factorizes into $AdS_5 \times \S^5/\Gamma$.
Only ${\mathcal N}=1$ SUSY is preserved
if the branes are placed at a conifold singularity
\cite{KW, MoPle, Acharya}.  The gauge theory
here is $SU(N) \times SU(N)$ with bifundamental matter, 
while the geometry
is $AdS_5 \times  T^{1,1}$

Unfortunately, $S^5$, $T^{1,1}$, and their orbifolds just about
exhaust the mathematical literature of 
SUSY preserving five dimensional Einstein spaces for
which explicit metrics are known.  In these cases, it 
was explicit knowledge of the metric which allowed for many tests of the 
AdS/CFT correspondence.
Nevertheless, there are an infinite number of five dimensional
$\bX$ from which one can construct AdS/CFT correspondences,
and it would be very useful to have techniques for dealing with
more general $\bX$.

In constructing these more general correspondences, we start
with a non-compact Calabi-Yau cone $\Y$, whose base is
the five dimensional Sasaki-Einstein manifold $\bX$.  Comparing
the metric with the D-brane description leads one to conjecture
that type IIB string theory on $AdS_5 \times \bX$ is dual to
the low-energy description of the worldvolume theory on the
D3-branes at the conical singularity \cite{MoPle, Acharya}.
It is known that for a stack of 
D3-branes placed at the conical singularity 
of $\Y$, the ten dimensional supergravity solution is
\[
ds^2 = h(r)^{-1/2} (-dt^2 + dx_1^2 + dx_2^2 + dx_3^2) + 
h(r)^{1/2} (dr^2 + r^2 ds_{\bX}^2) \ ,
\]
\be
h(r) = 1+ \frac{L^4}{r^4} \; \; , \; \; \; \;
L^4 = \frac{4\pi^4 g_s N \alpha'^2}{\Vol(\bX)} \ ,
\label{sugra}
\ee
\[
F_5 = {\cal F}_5 + \star {\cal F}_5 \ , \; \; \; 
{\cal F}_5 = 16\pi \alpha'^2 N \frac{\Vol(\S^5)}{\Vol(\bX)} \vol(\bX) \ ,
\]
where all the other field strengths vanish and $N$ is the number of D3-branes.
The constants $g_s$ and $\alpha'$ are the string coupling and the
Regge slope.
With this notation, $\vol$ is the volume differential form.  Thus
\[
\int_{\bX} \vol(\bX) = \Vol(\bX) \ .
\]
To get the space to factorize into $AdS_5 \times \bX$, 
we take the near horizon limit $r \ll L$.

As can be seen from (\ref{sugra}), it would be good to have a metric
independent expression for the $\Vol(\bX)$.  Indeed, in \cite{BH},
such an expression was derived for a large class of Sasaki-Einstein
manifolds $\bX$.  This volume is also important for calculating
the central charge $c$ of the gauge theory, as was pointed out
in \cite{Gubser, Kostas}.  In this paper, we will calculate the 
minimal volumes of three-cycles in $\bX$ in a metric
independent way.  These three-cycles are
important for calculating the conformal dimension of dibaryonic
operators and may have other uses as well.

In order to demonstrate that the AdS/CFT correspondence
is a duality between string theory and gauge theory
and not just between a supergravity theory and a gauge
theory, objects known as dibaryons and giant gravitons
have played an important role \cite{EW2, SMT}.  
On the string theory
side, dibaryons and giant gravitons correspond to 
supersymmetric D3-branes wrapped on three cycles 
${\mathcal H} \subset \bX$.
Roughly speaking, the mass $m$ of these D3-branes is 
just the D3-brane tension $\tau$ times
$\Vol({\mathcal H})$.
In fact, the mass suffers additional
zero mode corrections \cite{BHK}, 
and instead the conformal
dimension $\Delta$ is exactly
\be
\Delta = L^4 \Vol({\mathcal H}) \tau \ .
\ee
Geometrically, the conformal dimension
is the eigenvalue of the $r \partial / \partial r$
operator acting on a wrapped D3-brane state.
On the gauge theory side, these dibaryons 
correspond to antisymmetrized products of order $N$
matter fields.  The conformal dimension $\Delta$ 
of the dibaryon
is the number of bifundamental matter fields
in the antisymmetrized product multiplied by the
conformal dimension of the individual bifundamental
matter fields.   

An important piece of evidence that
dibaryons are alternately described either as wrapped
D3-branes or as antisymmetric products of order
$N$ matter fields is that the conformal dimensions,
when calculated in these two different ways,
agree.  The dibaryon dimension is calculated
from wrapped D3-branes at strong 't Hooft coupling
while the gauge theory calculation is naturally
a weak coupling result.  One may ask why these
conformal dimensions are not a function of
the 't Hooft coupling.  The reason is that
the dibaryons are BPS objects and their conformal
dimension is protected by the SUSY algebra
\cite{EW2}.

To understand the relationship between 
dibaryons and holomorphic curves mentioned in 
the abstract, we note that generically 
these Sasaki-Einstein manifolds 
$\bX$ can be expressed as $U(1)$ fibrations of
K\"{a}hler-Einstein surfaces $\bV$.  Similarly,
for the simplest, time independent dibaryons,
the wrapped three-cycles ${\mathcal H}$ 
correspond to $U(1)$ fibrations over holomorphic
curves $C \subset \bV$.  The K\"{a}hler-Einstein
relation allows us to relate $\Vol({\mathcal H})$
to the degree $-K_\bV \cdot C$ of the curve $C$.
Moreover, there should be a relationship between 
the number of curves $C$ of a given degree
and the number of dibaryons of a given conformal
dimension, which we will only begin
to explore in this paper.

We begin by giving a precise geometric description of
dibaryonic operators and derive a metric independent
formula for the volumes of the wrapped three-cycles.  Next,
we show that our formula gives the correct dimension
for giant gravitons in $\S^5$ and dibaryons in 
$T^{1,1}$ and $\S^5/\Z_3$.  Previous derivations
\cite{SMT,KG}
of these dimensions from geometry 
relied on an explicit knowledge
of the $\S^5$ and $T^{1,1}$ metrics.  

In the second half of the paper, we apply our formula
to study dibaryons in new contexts.
We apply our formula to 
and do some simple counting of 
dibaryons in U(1) bundles
over smooth del Pezzo surfaces.  Previously,
only dibaryons in the third del Pezzo had been
studied in detail \cite{Beasley1}.  
These U(1) bundles over smooth del Pezzo surfaces
are examples of Sasaki-Einstein manifolds $\bX$ and have
been heavily studied in the context of AdS/CFT
correspondence
\cite{Hanetal2, unify, IqHan, Hanetal}. 
We take advantage of some recent
progress in understanding the gauge theory
duals of these manifolds \cite{Wijnholt, IW}. 

Finally, 
we consider the dibaryons in the generalized
conifolds of Gubser, Nekrasov, and Shatashvili
\cite{GNS}.  These generalized conifolds can
be understood alternately as generalizations
of the $AdS_5 \times T^{1,1}$ correspondence
or as deformations of the $AdS_5 \times \S^5/\Gamma$
correspondences reviewed above.
In order to apply our volume formula, we will need
to develop some additional tools from algebraic
geometry to deal with the quotient singularities that appear
in studying these $\bX$.  For the dibaryons of smallest
conformal dimension, we are able to match
the number of holomorphic curves to the number
of gauge theory dibaryons for some of the 
generalized conifolds.

\section{The Geometry of the Dibaryon}

To understand these wrapped D3-branes better, let
us review an argument due to Mikhailov \cite{Mik} and
Beasley \cite{Beasley2} that relates the D3-brane
wrapping in $\bX$ to holomorphic four cycles in
the full Calabi-Yau cone $\Y$.  
We begin by thinking about Euclidean signature
ten dimensional space rather than Minkowski 
signature.
It is well
known that in the standard compactification of
ten dimensional space on a six dimensional Calabi-Yau
manifold, D-branes that wrap holomorphic four cycles
in $\Y$ preserve supersymmetry and hence will
be stable.   
When we add the $F_5$ flux from the $N$ 
D3-branes, the near
horizon geometry factorizes into
$\bX \times H^5$ where $H^5$ is 
Euclidean signature hyperbolic space.  
While the D3-brane looked like a point in
the Euclidean $\R{4}$, it now looks
like a line in $H^5$.  
The next step is to Wick rotate.  Wick rotation
will preserve the supersymmetry, and 
the
hyperbolic space $H^5$ is highly symmetric so
we can Wick rotate a coordinate that
will cause the D3-brane's path in $H^5$ to
be time-like in the resulting $AdS_5$.

From this construction of a giant graviton
or dibaryon, it is clear that the time 
coordinate in $AdS_5$ will be paired
holomorphically to a second coordinate
in $\bX$ to make what used to be a  
single complex coordinate in the cone
$\Y$.  
To see how this pairing
works, we need to investigate the
structure of $\bX$ in greater detail. 

Let $\bX$ be a Riemannian manifold of real dimension
$2n+1$
and $g_{ab}$ the associated Riemannian metric.
$\bX$ is defined to be Sasaki if the
holonomy group of the cone $\Y \equiv \R{+} \times \bX$ with metric
\be
ds^2_{\Y} = dr^2 + r^2 g_{ab} dx^a dx^b
\ee 
reduces to a subgroup of $U(n+1)$.  
Moreover, $\bX$ is Sasaki-Einstein if and only if
the cone $\Y$ is Calabi-Yau.
Let us specialize to the case where $\Y$ is
three complex dimensional.  One finds
$R_{ab} = 4 g_{ab}$ where $R_{ab}$ is the
Ricci tensor on $\bX$.

There is a natural $U(1)$ group action on $\bX$, and
a number of ways to understand where it comes from.
From AdS/CFT correspondence, we know the gauge
theories are conformal and have ${\mathcal N}=1$
supersymmetry, provided $\Y$ is Calabi-Yau.  Thus,
there will be a $U(1)$ R-symmetry.  Alternately,
there is a 
complex structure on $\Y$.  By taking the level surfaces
$\bX$ of $\Y$, we have essentially quotiented by the
absolute value of one of the complex coordinates on
$\Y$.  This complex coordinate also has a phase, which
corresponds to this $U(1)$ action. 

There exist a large class of $\bX$ called quasiregular
Sasaki-Einstein manifolds for which the orbits
of the $U(1)$ action are compact \cite{BG2}.   If $\bX$ is 
quasiregular, then it can be expressed as a circle bundle
$\pi : \bX \rightarrow \bV$ where $\bV$ is 
a K\"{a}hler-Einstein orbifold.  
Let $\pi^*$ be the pullback from $\bV$ to $\bX$.
Let 
\be
\omega = ih_{a\bar{b}} dz^a \wedge d\bar{z}^{\bar b}
\ee
be the K\"{a}hler form on $\bV$.  Define a one-form
$\eta$ on the fibration with curvature 
$d\eta = 2 \pi^* \omega$.  The metric
on $\bX$ is
\be
\bg = \pi^* \bh + \eta \otimes \eta \ .
\ee

If we let $\psi$ be an angular coordinate on the circle
bundle, this construction has hopefully made clear
that the radius $r$ and $\psi$ get paired holomorphically
as a single complex coordinate in $\Y$.  From our
construction of a wrapped D3-brane, we saw that $r$
gets turned into the time coordinate $t$.  Thus
$t$ and $\psi$ are paired.  The embedding of
the D3-brane can depend on time only in the combination
$t + \psi$.  (The coordinate $t$ can be thought of as
global time on AdS.)

In this paper, we will restrict to dibaryon and
giant graviton configurations which are time
independent.  (For more on time dependence,
see \cite{Beasley2, BHK}.)  As a result,
the configurations must also be $\psi$
independent.  In other words, the wrappings
will be invariant under the U(1) action 
of the fiber bundle.  The way to insure
this invariance is to make sure that
for every point $p \in \bV$
where the D3-brane is present, the
D3-brane also wraps completely the U(1)
fiber at that point.   Holomorphicity
then requires that these points 
$p \in \bV$ trace out a holomorphic
curve in $\bV$ which we shall call $C$.

Naively, one might have thought our D3-brane
wrappings would correspond to topologically 
distinct three-cycles of $\bX$.  This
intuition is not correct.  There are
many more holomorphic curves $C$ 
in $\bV$ then
there are homology cycles although in
general we can build $C$ out of 
homology cycles in $\bV$.  The minimal
volume curves should correspond
to homology cycles in $\bV$, but 
they do not, always, correspond to
homology cycles in $\bX$.  Instead,
the minimal volume curves 
correspond to equivariant
homology cycles in $\bX$.

Let $\omega$ be the
K\"{a}hler form on $\bV$.  Let 
$\omega_i$, $i=1,\ldots,n$ correspond to
the other harmonic forms on $\bV$.  
(It is possible there are no harmonic
forms besides $\omega$.)  
We can insist
that $\omega \wedge \omega_i = 0$.
To construct harmonic forms on $\bX$,
we wedge with the one form $\eta$.
The set of harmonic three forms on $\bX$
is given by 
$\{\eta \wedge \pi^* \omega_i : i=1,\ldots n\}$
where $\pi^*$ is the pull back map of the
fibration.
Note that $\eta\wedge \pi^* \omega$ is not closed,
and so the Betti number of $\bX$ $b_3=n$ will
be one less than the Hodge number $h^{1,1}=n+1$
of $\bV$.  

It turns out, however, that the dimension
of the equivariant homology is
exactly the dimension of $H^{1,1}(\bV)$.
Consider the two-form $\pi^* \omega$.  It is
not harmonic because $d\eta = 2 \pi^* \omega$.
However, we can't use $\eta$ in calculating
the equivariant cohomology because $\eta$
is not invariant under the $U(1)$ action.
A more careful analysis shows 
that $\omega$ is a good representative
of the equivariant cohomology, and so its dual
cycle is a good representative of the equivariant
homology.

Now to compute volumes of these submanifolds in $\bX$,
we shall make use of the Einstein condition.
In particular, the fact that 
$\bV$ is K\"{a}hler-Einstein means
$\bR = 6 \bh$, where $\bR$ is the Ricci tensor
on $\bV$.  
The first Chern class of $\bV$, denoted $c_1(\bV)$,
is given locally by
\be
c_1(\bV) = i \frac{R_{a\bar{b}}}{2\pi} 
dz^a \wedge d\bar{z}^{\bar b} \ .
\ee
The critical step here is to use the K\"{a}hler-Einstein
condition to write the K\"{a}hler form $\omega$ in terms
of the first Chern class
\be
\omega = \frac{\pi}{3} c_1 (\bV) \ .
\ee
It is this relation which obviates the need for explicit
knowledge of the metric on $\bV$.  Volumes, which
are proportional to integrals over $\omega$,
can now be expressed as integrals over
the curvature, which usually have a topological
interpretation.

Now the volume of $\bV$, as was found in \cite{BH} is
(see also \cite{Besse})
\be
\Vol(\bV) = \frac{1}{2} \int_\bV \omega \wedge \omega =
\frac{\pi^2}{18} \int_\bV c_1(\bV)^2 \ .
\ee
If we define $K_\bV$ to be the canonical line 
bundle over $\bV$, one finds
\be
\Vol(\bV) = \frac{\pi^2}{18} K_\bV \cdot K_\bV \ .
\label{volV}
\ee 
In other words, the volume of $\bV$ is related in a
simple way to the self-intersection number of the
canonical line bundle.

The quasiregular condition means additionally that the length of the U(1)
fiber does not vary as we move around in $\bV$.  Thus,
the volume of the manifold $\bX$ is the volume of the 
manifold $\bV$ times the length of the U(1) fiber.  Similarly,
the volume of the three-cycle ${\mathcal H}$ will be the area of the
corresponding holomorphic curve $C$ in $\bV$ times the length of
the U(1) fiber.

To make the gauge theory comparison, we need to calculate
the conformal dimension 
of a D3-brane wrapped on such a cycle ${\mathcal H}$.
As was shown in \cite{BHK} and mentioned in the introduction
\[
\Delta = L^4 \Vol({\mathcal H}) \tau \ ,
\]
where $\tau = 1/ 8 \pi^3 g_s \alpha'^2$ is the D3-brane
tension.  Then from the quantization condition 
(\ref{sugra}) on $L$
\be
\Delta = L^4 \Vol({\mathcal H}) \tau = \frac{\pi N}{2}
\frac{\Vol({\mathcal H})}{\Vol(\bX)} \ .
\ee
Both $\Vol({\mathcal H})$ and $\Vol(\bX)$ include the same factor of
the length of the U(1) fiber.  We divide out to find
\be
\Delta = \frac{\pi N}{2} \frac{\Vol(C)}{\Vol(\bV)} \ .
\label{mLone}
\ee

We can relate the $\Vol(C)$ to an intersection number calculation.  
In particular
\be
\Vol(C) = \int_C \omega = \frac{\pi}{3} \int_C c_1(\bV) =
-\frac{\pi}{3} K_{\bV} \cdot C \ .
\label{volC}
\ee
Putting the pieces 
(\ref{volV}), (\ref{mLone}), and (\ref{volC})
together, we arrive at our final
formula for the dibaryon dimension
\be
\Delta = - 3N \frac{K_{\bV} \cdot C}{K_{\bV} \cdot K_{\bV}} \ .
\label{result}
\ee
For smooth manifolds $\bV$, the intersection numbers 
$K_\bV \cdot K_\bV$ and $K_\bV \cdot C$ are 
integers.  When $\bV$ is an orbifold however, these 
intersections are generally rational numbers.

\section{Dibaryons in $T^{1,1}$ and Giant Gravitons in $\S^5$}

This formula (\ref{result}) is rather powerful.  To understand how to
apply it, let us begin with examples of $\bV$ where
explicit metrics are known and where the calculation
of $\Delta$ was performed originally in a more straightforward
manner.

\subsection{$AdS_5 \times \S^5$}

We begin with $AdS_5 \times \S^5$.  The sphere
$\S^5$ can be thought of as a U(1) bundle over
$\CP{2}$.  The surface $\bV = \CP{2}$ is clearly a
K\"{a}hler-Einstein space.  Let $H$ be the
hyperplane bundle on $\CP{2}$.  The canonical
bundle on $\CP{2}$ is $K_{\CP{2}} = -3H$.  
The intersection $H\cdot H = 1$.  This 
relation is just a formal statement of the
fact that two lines intersect at a point.
Let our holomorphic curve be $H$.  
From (\ref{result}),
one finds that $\Delta = N$ for the choice 
$C = H$.

This result $\Delta= N$ for $AdS_5 \times \S^5$ has
two different possible interpretations.
The first interpretation involves the
giant gravitons of \cite{SMT}.
The gauge theory dual of $AdS_5 \times \S^5$
is ${\mathcal N}=4$ $SU(N)$ super Yang Mills, 
which contains three scalars, $X$, $Y$, and $Z$, 
transforming in the
adjoint of $SU(N)$.  Each of these scalars
has conformal dimension 1.  There is clearly
a set of operators of dimension $N$ 
obtained by antisymmetrizing over a product
of $N$ of the $X$'s, $Y$'s, and $Z$'s
\be
\epsilon^{i_i i_2 \ldots i_N}
\epsilon_{j_1 j_2 \ldots j_N}
X^{j_1}_{i_1} \cdots X^{j_x}_{i_x} 
Y^{j_{x+1}}_{i_{x+1}} \cdots 
Y^{j_{y}}_{i_{y}} Z^{j_{y+1}}_{i_{y+1}}
\cdots
Z^{j_N}_{i_N} 
\ee
where $x\leq y$ are some integers.
These operators are the maximal giant gravitons
of \cite{SMT}.  They are not topologically
stable but rather dynamically stabilized.
So despite the fact that it doesn't make sense
to think about D3-branes wrapping topologically
nontrivial cycles in $\S^5$, the topology
of the underlying $\CP{2}$ is
all that's needed to understand the 
conformal dimension of the
maximal giant gravitons.  There also
exist smaller giants where some of the
$X$, $Y$, and $Z$ are replaced with the
identity operator.  While the analysis
of \cite{SMT} shows that the maximal
giants are time independent, the 
sub-maximal giants spin in the transverse
$\S^5$, are thus time dependent,
and are beyond the scope of this paper.

The other interpretation involves the fact
that there are actually two distinct U(1)
fibrations over $\CP{2}$ that yield
Sasaki-Einstein spaces.  The naive fibration
produces $\S^5$.  However, if we shrink the
length of the U(1) fiber by a factor of three,
one gets the orbifold $\S^5/\Z_3$
which was well studied by \cite{Z3orb}.\footnote{
The different possible fibrations come from
the Thom-Gysin sequence which implies 
that the first Chern class
of the circle fibration $c_1^*$ 
divides $c_1(\bV)$
\cite{FrK1, FrK2}.  For $\S^5$, 
$c_1(\bV) = 3H$ and so $c_1^*$ is 
either $H$ or $3H$.}
This geometry has a dual gauge theory
with gauge group 
$SU(N)\times SU(N) \times SU(N)$.  
There are three sets
of three bifundamental matter fields $X_i$, $Y_i$,
and $Z_i$, $i=1$, $2$, or $3$, transforming in the bifundamental
representations of each pair of the three
$SU(N)$'s.  The conformal dimension of each of the 
bifundamental matter fields is still one as all we
have done is orbifold.
This
space $\S^5/\Z_3$ has nonvanishing homology class 
$H_3(\S^5/\Z_3) = \Z_3$,
and so it makes sense to speak of nontrivially wrapped D3-branes.
Indeed, $\Delta=N$ is the right prediction for an antisymmetric
product of $N$ of the bifundamental matter fields.

\subsection{$AdS_5 \times T^{1,1}$}

This 
calculation was done first by
Gubser and Klebanov \cite{KG}.  
We repeat their calculation with our 
advanced technology.
To produce the space $T^{1,1}$ 
from a U(1) fibration, one
takes the underlying K\"{a}hler-Einstein
space to be $\bV = \CP{1} \times \CP{1}$.
There is also a second U(1) fibration which results in 
$T^{1,1}/\Z_2$, but as the dimension $\Delta$ remains the 
same, we will focus on $T^{1,1}$ in what follows.

The canonical class $K_\bV = -2f -2g$ where $f{\cdot}g = 1$,
$f\cdot f=0$, and $g \cdot g = 0$.  One can think of the
line bundles $f$ and $g$ as corresponding to the 
individual $\CP{1}$'s.  Each $\CP{1}$ does not intersect
with itself and intersects with the other $\CP{1}$ exactly
once.  The simplest holomorphic curve one can take is
$C=f$ or equivalently $C=g$.  Based on this construction,
one sees easily that $\Delta = 3N/4$.

The gauge theory dual to $AdS_5 \times T^{1,1}$ is
${\mathcal N}=1$ $SU(N)\times SU(N)$
with two bifundamental matter fields $A$ and $B$.
The $A$ fields transform under $({\mathbf N}, 
\bar{\mathbf N})$ while the $B$ fields transform
under $(\bar{\mathbf N}, {\mathbf N})$.  
A dibaryon $D$ is a product of $N$ $A$'s or alternately
of $N$ $B$'s, totally antisymmetrized with 
respect to both color indices:
\begin{equation}
D = \epsilon^{\alpha_1 \alpha_2 \cdots \alpha_N}
\epsilon_{\beta_1 \beta_2 \cdots \beta_N}
A_{\alpha_1}^{\beta_1} A_{\alpha_2}^{\beta_2} \cdots
A_{\alpha_N}^{\beta_N} \ .
\end{equation}
The conformal dimension of the $A$ and $B$ fields is
3/4.  It follows that the total dimension 
$\Delta(D)$ is $3N/4$.  

\section{Smooth $\bV$} 

We now consider the case when $\bV$ is a 
K\"{a}hler-Einstein manifold without
orbifold singularities.  In addition
to $\CP{2}$ and $\CP{1}\times \CP{1}$
considered above, there are remarkably
few such spaces.  They are the
del Pezzo surfaces where $\CP{2}$ has
been blown up at $n$ points,
$3 \leq n \leq 8$.\footnote{
No three points should be collinear and
no six points should lie on a conic.
}  
Let $H$ be the 
hyperplane bundle in $\CP{2}$.
The canonical class of the $n$th
del Pezzo surface is
\be
K_n = -3H + \sum_{i=1}^n E_i
\ee
where $E_i$ is the exceptional 
divisor of a blown up point.
Again $H\cdot H = 1$.  Moreover,
$E_i \cdot E_j = - \delta_{ij}$, 
and $H \cdot E_i = 0$.

In general, we can pick
$C$ such that $-K_\bV \cdot C$ is any
positive integer $k$.  
From our construction,
$K_n \cdot K_n = 9-n$.  Thus
\be
\Delta = k\frac{3N}{9-n} \ .
\label{dpresult}
\ee
As an example, take the line $C = E_i$ as our
holomorphic curve inside $\bV$.  
Thus 
$K_n \cdot E_i = -1$, and the curve
$C$ has $k=1$.

In addition to matching the spectrum
$\Delta$ of conformal dimensions from gauge theory, 
we can also do some elementary counting of dibaryons.
In preparation for this counting, note that  
the curve $C$ can be written 
\be
C = a H - \sum_{i=1}^n b_i E_i
\ee
where the $a$ and $b_i$ are integers.  The genus
formula tells us that
\be
(K_n + C ) \cdot C = 2g - 2
\ee
where $g$ is the genus of the curve.  Moreover,
$g\geq 0$.  
Let us count how many degree $k=1$ curves there are for each
del Pezzo, taking into account the constraint $g \geq 0$.
It turns out that the only $k=1$ curves also have $g=0$
(except for the eighth del Pezzo which also has a
$k=1$ curve of $g=1$ corresonding to $-K_8$):
\be
\renewcommand{\arraystretch}{1.2}
\begin{array}{ccccccccc}
n & 1 & 2 & 3 & 4 & 5 & 6 & 7 & 8 \\
\mbox{\# of curves} &
1 & 3 & 6 & 10 & 16 & 27 & 56 & 241 
\end{array}
\label{counting}
\ee

The gauge theory duals of the del Pezzos are more
complicated because of a phenomenon known variously
as Seiberg duality, toric duality, and Picard-Lefschetz
monodromy \cite{unify, Beasley1, Hanetal2}.  In simple
words, a single Calabi-Yau cone over a del Pezzo has 
more than one gauge theory dual description.
In fact, there are infinite 
families of gauge theories which all correspond
to the same cone over a del Pezzo!  
In the following, we will for the most part content
ourselves with looking at a single one of the possible
gauge theory duals for each del Pezzo.  

\subsection{The Third del Pezzo}

The third del Pezzo, $n=3$, was studied in great
detail by Beasley and Plesser \cite{Beasley1}.
For completeness, we give a brief summary
of their discussion of dibaryon dimensions.
Beasley and Plesser studied four of the 
Seiberg dual
gauge theories which map to this
third del Pezzo through AdS/CFT correspondence.
These four gauge theories have complicated
chiral quiver diagrams involving six gauge groups
and a large number of bifundamental matter fields.

To check the calculation of $\Delta$, we need not be
concerned by these complications.  Beasley
and Plesser tell us that the bifundamental
matter fields present in the four
quiver theories they considered 
all have conformal dimension $1/2$, 1, or $3/2$.  
Moreover, the gauge groups are all $SU(N)$.  In 
other words, $\Delta$ should be an integer multiple of $N/2$.
To get $N/2$, we choose for example $C=-E_i$. 
To get these larger dimensions, we need to
choose a different holomorphic curve.
For example, $C = H - E_i$ yields $\Delta = N$, and
$C = 2H - E_1 - E_2 - E_3$ yields $\Delta = 3N/2$.

As \cite{Beasley1} did before us, 
we may also count the number of dibaryons with the smallest
conformal dimension $N/2$.  In each of the four quiver 
theories considered, there are six bifundamental matter
fields with conformal dimension $1/2$, corresponding
to the six degree one curves in table
(\ref{counting}).  

The authors of \cite{Beasley1} also count the dibaryons of
dimensions $N$ and $3N/2$.  Counting higher degree curves
is easy from the geometric point of view.  However,
the gauge theory story is
more complicated.
The naive number of these 
dibaryons from gauge theory is far too large and
gets reduced by classical and ``quantum'' relations between
the bifundamental matter fields.  We refer the reader to
\cite{Beasley1} for a discussion of these counting complications for
the third del Pezzo.

\subsection{del Pezzos Five and Six}

The gauge theory duals of the 
fifth and sixth del Pezzos were considered
by \cite{IqHan, Hanetal, Wijnholt}.  
The conformal dimension of the dibaryon operators
in both cases can be understood from
the $AdS_5 \times \S^5$ and $AdS_5 \times T^{1,1}$
examples studied above.
There are quivers for the fifth and sixth
del Pezzo which are identical to the quivers
for orbifolds of $T^{1,1}$ and $\S^5$.
In particular, there is
a quiver for the fifth del Pezzo 
which is identical to the quiver for a $\Z_2 \times \Z_2$
orbifold of $T^{1,1}$.  
The smallest dibaryons for the fifth del Pezzo 
have dimension $3N/4$,
just like the $T^{1,1}$ case.
There is a quiver for the sixth del Pezzo which
is identical to the quiver for
a  $\Z_3 \times \Z_3$ orbifold of $\S^5$.
Not surprisingly, we find that the smallest
dibaryon for the sixth del Pezzo 
has dimension $N$, just like
the $\S^5$ case.

Geometrically, one can understand roughly how
these relations to orbifold quivers arise.
There are limits of the 
del Pezzo surfaces where three of the blown up points
lie on a line or six lie on a conic.
In these limits, the surface may admit a simpler 
orbifold description.  We know there is
a limit of the fifth del Pezzo which gives rise
to this $\Z_{2} \times \Z_{2}$ orbifold and a limit
of the cone over the sixth del Pezzo which gives
$\C{3} / \Z_{3} \times \Z_{3}$. 

We can count the number of smallest dibaryons.
The quiver for the fifth del Pezzo (see figure \ref{figdp4})
has 16 lines, corresponding to 16 bifundamental matter
fields with conformal dimension $3/4$.  From table
(\ref{counting}), we see that there are exactly 16 
degree one curves in the fifth del Pezzo.
Similarly, for the sixth del Pezzo, there are 27 bifundamentals
with conformal dimension 1, corresponding both to the 27
degree one curves in the sixth del Pezzo and the 27
lines in the quiver of figure \ref{figdp4}.

\subsection{The Fourth del Pezzo}

\begin{figure}
\includegraphics[width=3.5in]{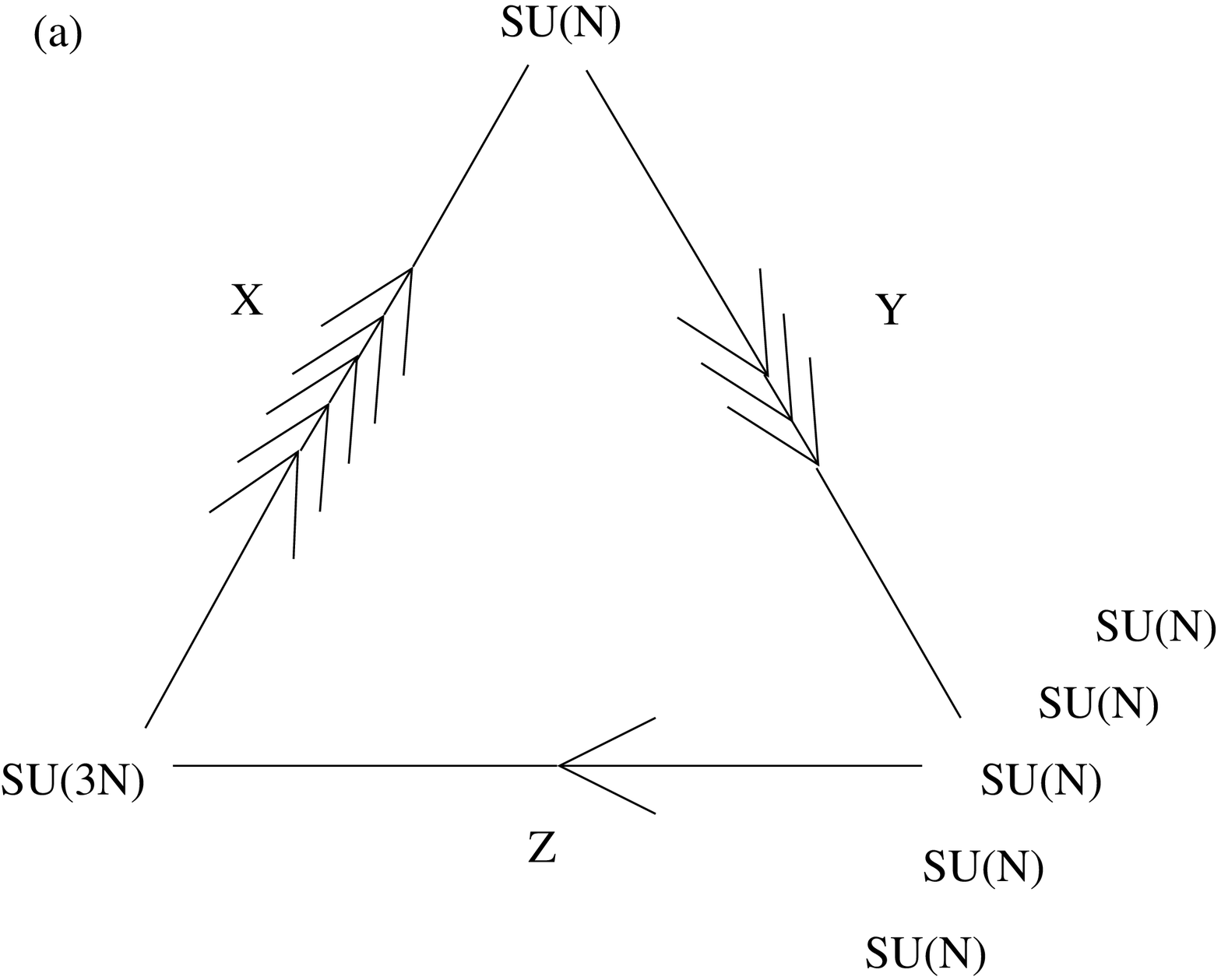}
\includegraphics[width=3.5in]{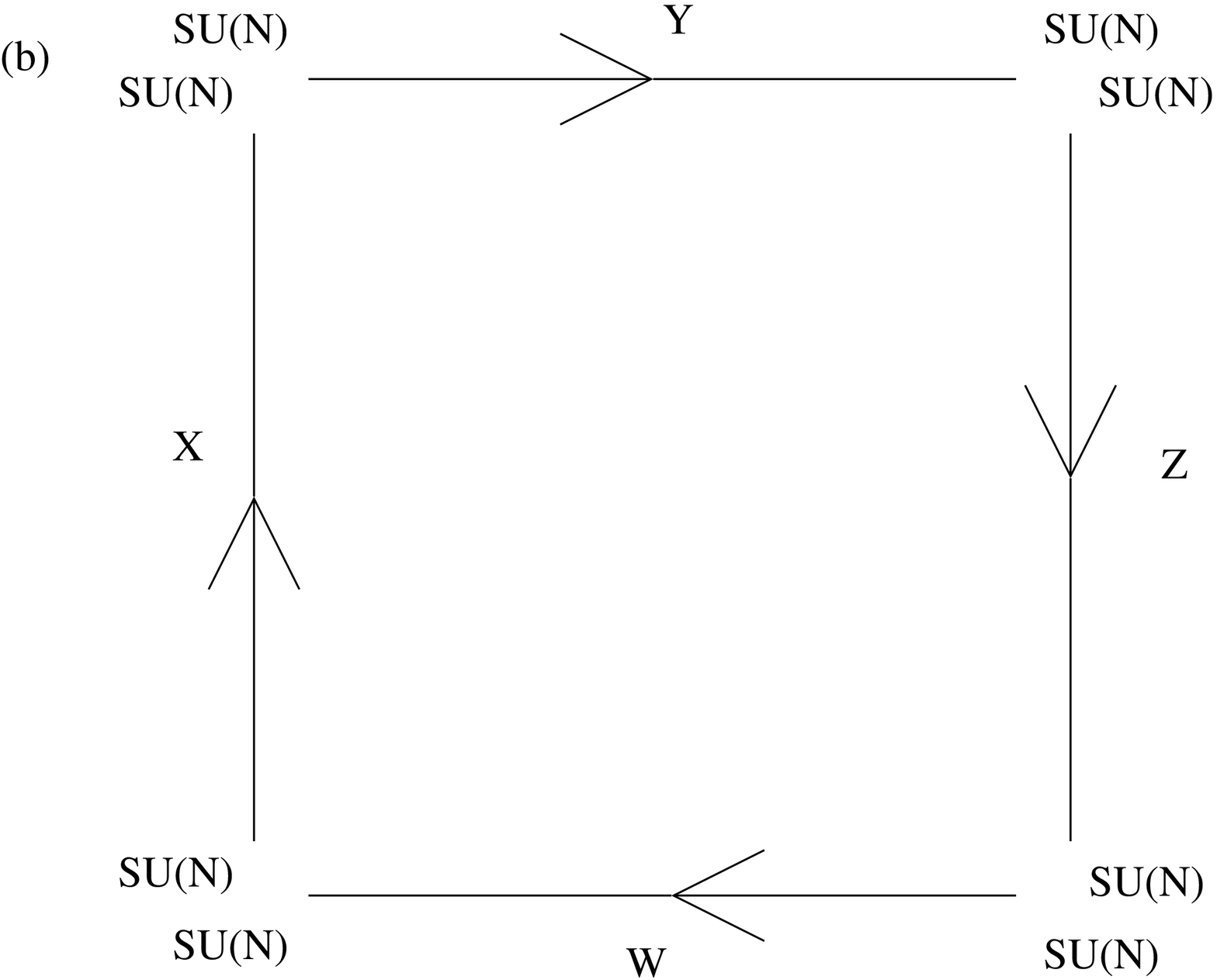}
\vfil
\vfil
\begin{center}
\includegraphics[width=3.5in]{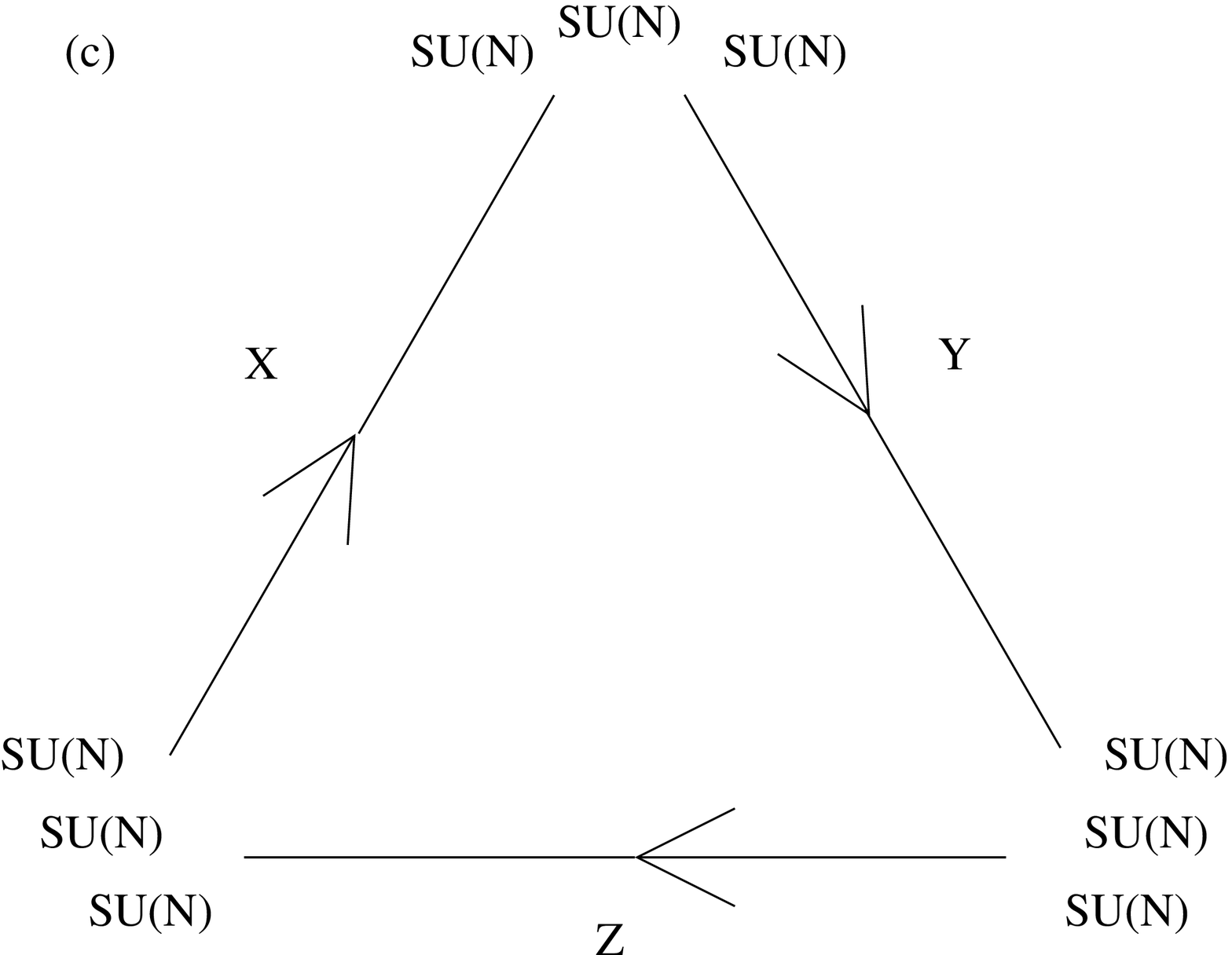}
\end{center}
\caption{Quivers of \cite{Wijnholt}
for the (a) fourth, (b) fifth, and (c) sixth del Pezzos.  In this 
condensed notation, each $SU(N)$ represents
a node.  For example, for the fourth
del Pezzo, a pair of bifundamentals
$Y$ and $Z$ attaches to each $SU(N)$ node in the
lower right hand corner of the quiver.}
\label{figdp4}
\end{figure}

A less trivial example is the fourth del Pezzo.
Wijnholt has produced a quiver and
superpotential for the gauge theory dual 
(Eq. 3.15 of \cite{Wijnholt}).  
There are five fields
$X_i$ transforming under the bifundamental of
$SU(3N) \times SU(N)$, five fields
$Z_i$ transforming in the bifundamental of 
$SU(N) \times SU(3N)$, and fifteen fields $Y_i$
transforming in the bifundamental of $SU(N) \times SU(N)$,
as shown in figure \ref{figdp4}.
The superpotential is constructed out of sums of the
$\Tr (X_i Y_j Z_k)$.  Vanishing of the beta functions and the fact that
the superpotential has R-charge two 
completely specify 
the conformal dimension of the bifundamental matter
fields.    
In particular, from the supersymmetry algebra, we know
that the conformal dimension of the matter fields is
$3/2$ their R-charge.  Thus from the superpotential
constraint, it follows that 
\be
\Delta_X + \Delta_Y + \Delta_Z = 3 \ .
\ee
For each node, we get an additional constraint from the 
vanishing of the NSVZ beta function:
\be
0 = 3 C_2(G) - \sum_i T(R_i) (3-2\Delta_i) \ ,
\ee
where $C_2(G)$ is the Casimir of the gauge group and
$T(R_i)$ is the index of representation of the matter
field.  For $SU(N)$, $C_2 = N$ and the index of the fundamental
representation is $1/2$.  From these constraints,
we learn that the $X$ have conformal
dimension 1, the $Y$ have conformal dimension 9/5, and the
$Z$ have conformal dimension 1/5.

This quiver for the fourth del Pezzo is the first example we have come across
where the gauge groups 
are not all $SU(N)$, and
we have to be a little careful about constructing
dibaryons.  In general, if we have operators
$({\mathcal O}_i)^\alpha_\beta$ transforming in the 
bifundamental of $SU(aN) \times SU(bN)$ where $a$ 
and $b$ are integers, we will need $N \, \lcm(a,b)$
copies of ${\mathcal O}$ in order to be able to
antisymmetrize completely over both color indices.  
Moreover, for each antisymmetrization over $SU(aN)$
or $SU(bN)$, the $aN$ or $bN$ ${\mathcal O}_i$ need
to be distinct in some way.
In our case,
$a=1$ and $b=3$; the smallest dibaryon that can
be constructed from the $X_i$ looks like
\be
\epsilon^{\beta_1 \cdots \beta_N}
\epsilon^{\gamma_1 \cdots \gamma_N}
\epsilon^{\delta_1 \cdots \delta_N}
\epsilon_{\alpha_1 \cdots \alpha_{3N}}
X^{\alpha_1}_{\beta_1} \cdots X^{\alpha_N}_{\beta_N}
X^{\alpha_{N+1}}_{\gamma_1} \cdots X^{\alpha_{2N}}_{\gamma_N}
X^{\alpha_{2N+1}}_{\delta_1} \cdots X^{\alpha_{3N}}_{\delta_N}
\ee
and has conformal dimension $3N$.  The conformal dimensions
of the $Y$ and $Z$ dibaryons are correspondingly
$9N/5$ and $3N/5$.  All of these numbers are
integer multiples of $3N/5$ as predicted by
(\ref{dpresult}).  To get a dibaryon of dimension
$6N/5$, one could take the bifundamental field
$ZX$, where we have traced over the internal $SU(3N)$
color indices.  An antisymmetric product of 
$N$ copies of the $ZX$ 
does indeed have conformal dimension $6N/5$.

From table (\ref{counting}), there are 10 degree one curves in 
the fourth del Pezzo, and hence there should be 10 dibaryons
with conformal dimension $3N/5$.  Note that there are
five bifundamental fields $X_i$.  In constructing a dibaryon,
we get to choose any three of them, and five choose three is
indeed 10.  If we choose the same $X_i$ twice, the
antisymmetrization gives zero.

\subsection{The First and Second del Pezzos}

Despite the fact that the first and second del
Pezzo are not K\"{a}hler-Einstein, let us
try applying our formula (\ref{dpresult})
anyway.  
For the first and second del Pezzo, the vanishing
of the NSVZ beta functions and the R-charge
constraint from the superpotential are not
sufficient to specify completely the conformal
dimension of the bifundamental matter fields.
Intriligator and Wecht \cite{IW} proposed recently
an additional constraint on these conformal
dimensions.  They demonstrated that 
the conformal dimensions or equivalently the
R-charges should be chosen in a way that
maximizes the conformal anomaly $a$.
In particular, the conformal anomaly
$a$ is proportional
to
\be
a \sim 3 \sum_i r(\psi_i)^3 - \sum_i r(\psi_i) \ 
\ee
where $r(\psi_i)$ is the R-charge of the fermion
$\psi_i$ and the sum runs over all species 
of fermion in the gauge theory.

\begin{figure}
\begin{center}
\includegraphics[width=2.5in]{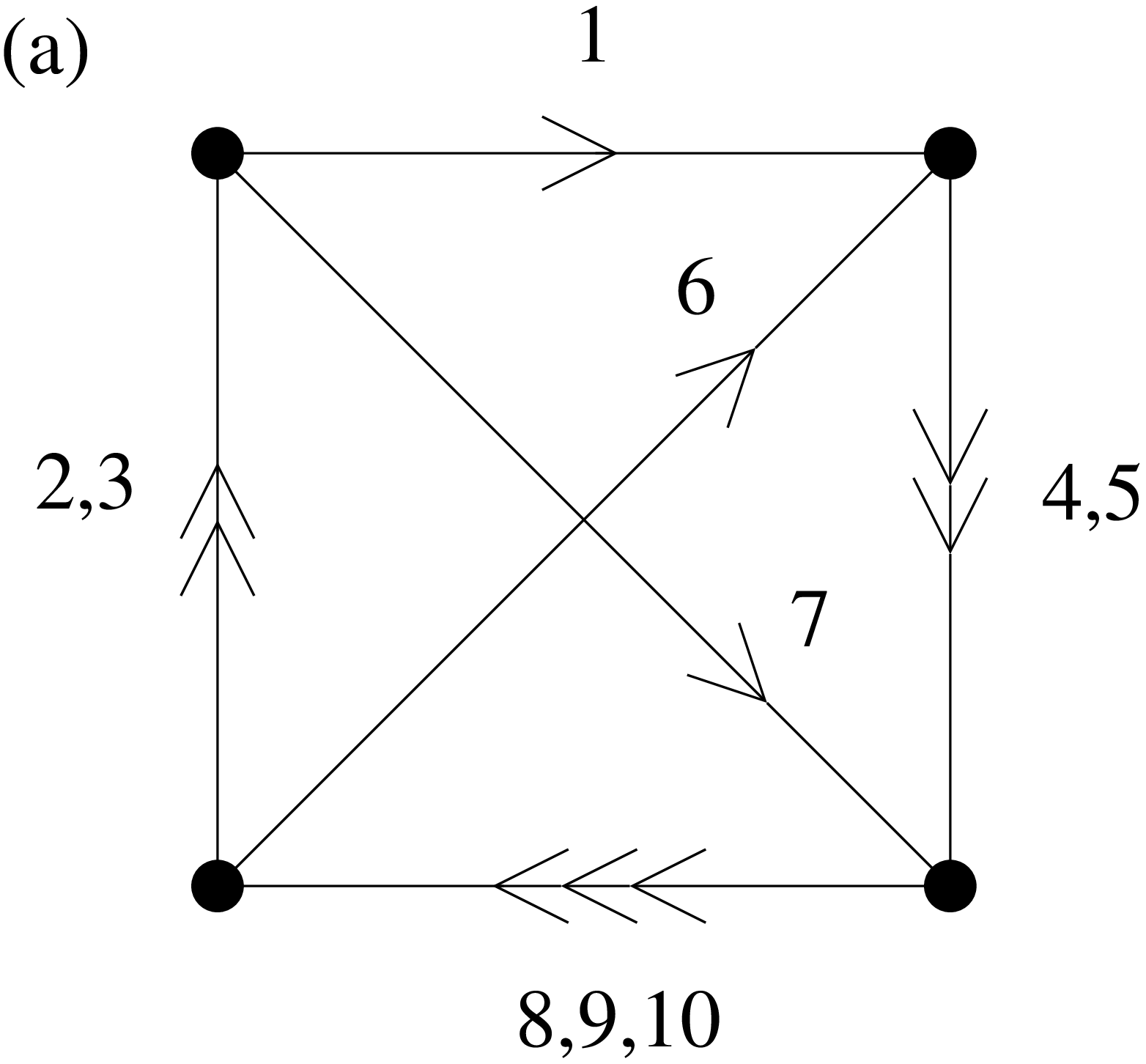}
\includegraphics[width=2.75in]{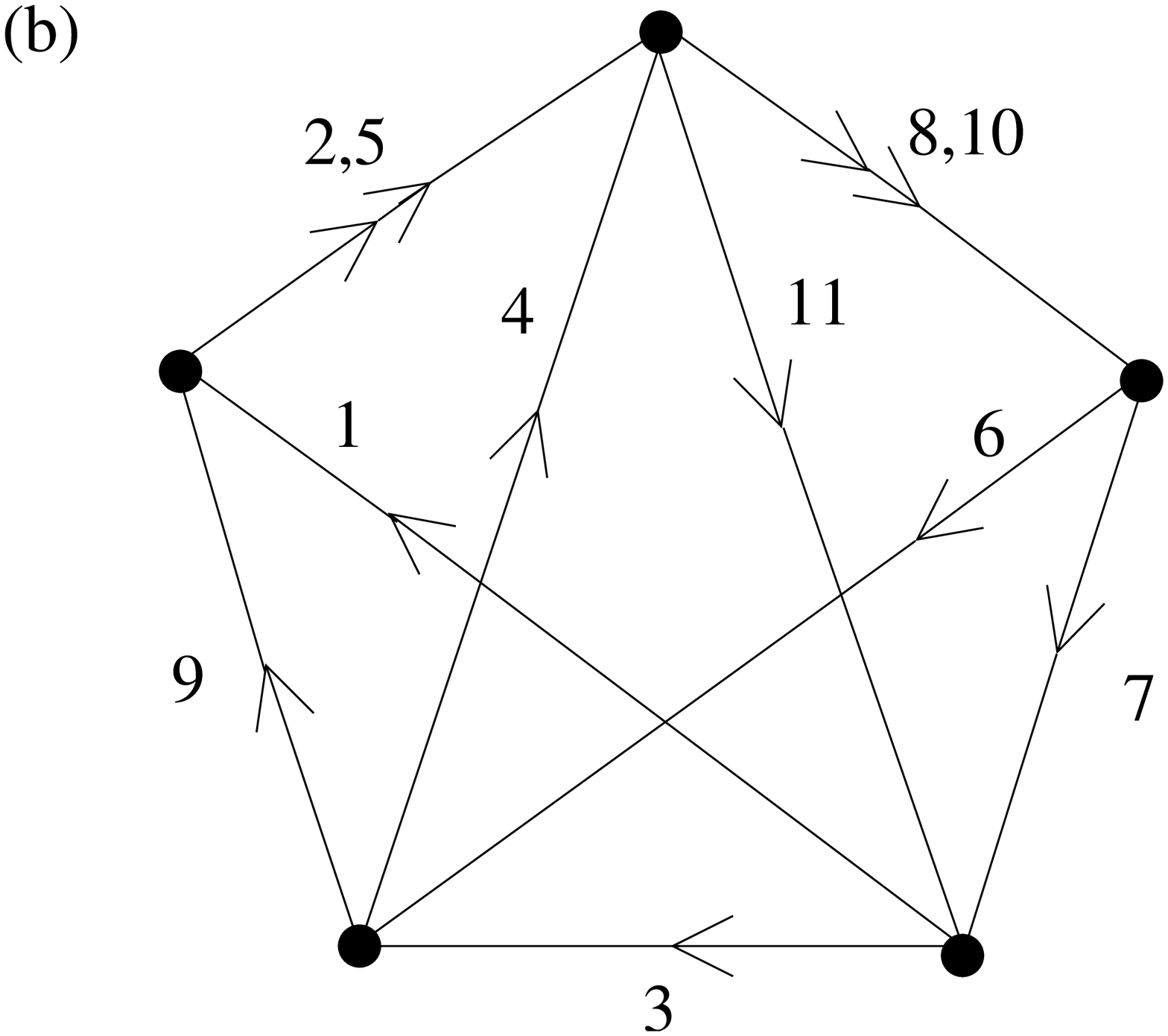}
\end{center}
\caption{a) The quiver of 
\cite{IW} for the first del Pezzo. 
b) The Model II quiver of \cite{Hanetal2}
for the second del Pezzo.  The nodes correspond
to $SU(N)$ gauge groups.}
\label{figdp2}
\end{figure}

Using this additional constraint,
Intriligator and Wecht \cite{IW}
have calculated
the conformal dimensions of the bifundamental
matter fields for a particular
gauge theory dual to the first
del Pezzo
(see figure \ref{figdp2}a).  The gauge groups are all
$SU(N)$, and there  are
10 bifundamental matter fields $X_i$,
$i=1,2,\ldots, 10$.
The allowed dimensions 
of these $X_i$ are
3/8, 3/4, and 9/8, as predicted by 
(\ref{dpresult}).  Moreover,
there is exactly one bifundamental matter
field, $X_1$, with conformal dimension 3/8,
corresponding to the single degree
one holomorphic curve for the first
del Pezzo in table (\ref{counting}).

The conformal dimensions of the bifundamental matter
fields have not yet been calculated for the 
second del Pezzo.  Let us take the quiver and
superpotential from Model II,
Eq. 4.4 of \cite{Hanetal2}.  
All the gauge groups in this model are $SU(N)$.
We have reproduced their quiver as figure 
\ref{figdp2}b.  As can be seen from the 
quiver, there are 11 bifundamental matter fields, 
$X_i$, $i=1,2,\ldots 11$.  The superpotential for
this theory is
\be
X_5 X_8 X_6 X_9 + X_1 X_2 X_{10} X_7 + X_{11} X_3 X_4
- X_4 X_{10} X_6 - X_2 X_8 X_7 X_3 X_9 - X_{11} X_1 X_5 \ .
\ee
The 
constraints from the vanishing of the
beta functions for each $SU(N)$ gauge group
and the fact that the superpotential
have R-charge 2 yield a two parameter family of
solutions for the dimensions
of the bifundamental matter fields.  
We then
maximize the conformal anomaly $a$,
as described by \cite{IW}.  
We find that
the conformal dimensions of
$X_3$, $X_7$, and $X_9$ are 3/7.
The conformal dimensions of
$X_1$, $X_2$, $X_5$, $X_6$, $X_8$, and
$X_{10}$ are all 6/7.  Finally, 
the conformal dimensions of $X_4$ and
$X_{11}$ are each 9/7.  These numbers,
3/7, 6/7, and 9/7, are exactly what
one expects from (\ref{dpresult}).
Moreover, there are three fields with
conformal dimension 3/7, as predicted
by table (\ref{counting}).

This calculation can be repeated for
the slightly more complicated Model I
gauge theory dual to the second
del Pezzo
of \cite{Hanetal2}.  The quiver and
superpotential of Model I are related
to Model II through a Seiberg duality
on one of the nodes of the quiver.
The gauge groups are all $SU(N)$.
There are thirteen matter fields instead
of eleven, but their conformal dimensions
follow the same sequence $3/7$, $6/7$,
$9/7, \ldots$ of rational numbers.  
Moreover, only three of the thirteen fields
have conformal dimension 3/7.
Thus, we find again the dibaryon
spectrum of (\ref{dpresult})
and table (\ref{counting}).

\subsection{del Pezzos Seven and Eight}

The cases $n=7,8$ are more troublesome because
the gauge theory descriptions are still incomplete.
Quivers but no superpotentials have been proposed
by \cite{IqHan}. 
We can make a prediction, however, for a quiver gauge theory
dual to the seventh or eighth del Pezzo.

Let us start with
the most troublesome case, $n=8$.  The geometric calculation
tells us that $\Delta$ is an integer multiple of $3N$.  If the gauge
groups in the quiver were all SU(N), we would conclude that the smallest
conformal dimension of a bifundamental matter field is 3 and thus
the smallest R-charge 2.  We could then conclude that the superpotential
vanishes because a loop in the quiver needs at least two bifundamental 
matter fields to close on itself.  We thus have two possibilities.
Either the superpotential vanishes or the gauge groups are not all $SU(N)$.

Let us assume for a 
moment that some of the gauge groups are not
$SU(N)$ but $SU(\alpha_i N)$ where the $\alpha_i$ are some
integers. As we discussed in the case of the fourth del Pezzo
surface, for a collection of bifundamental operators 
$({\mathcal O}_k)^\alpha_\beta$ transforming under
$SU(\alpha_i N) \times SU(\alpha_j N)$ one needs now 
$N \, \lcm(\alpha_i, \alpha_j)$ copies
of ${\mathcal O}_k$ to antisymmetrize properly.  For
the eighth del Pezzo, the minimum conformal dimension
of such an ${\mathcal O}_k$ is reduced from 3 to 
$3/\lcm(\alpha_i, \alpha_j)$.

For the case $n=7$, if we assume all the gauge
groups are SU(N), the minimal naive dimension of the
bifundamental matter fields is 3/2.
So for $n=7$, it might be possible to have a quadratic
superpotential.  However, the quiver in \cite{IqHan}
is chiral and has no loops with only two bifundamental
matter fields.  Thus it seems reasonable to 
conjecture that some of the gauge groups are not
pure $SU(N)$ but $SU(\alpha_i N)$ for $\alpha_i$ integer.

That exhausts the collection of smooth K\"{a}hler-Einstein
manifolds.  Next we turn to $\bV$ with quotient singularities.

\section{The Generalized Conifolds}

Consider a 
weighted homogenous polynomial in $\C{4}$, by which we mean a
polynomial $F(\bz)$ which satisfies
\[
F(\lambda^{w_0} z_0, \lambda^{w_1} z_1 , \lambda^{w_2} z_2 , 
\lambda^{w_3} z_3) =
\lambda^d F(z_0, z_1, z_2, z_3) \ ,
\] 
where $\lambda \in \C{*}$ and $w_i \in \Z^+$, and the 
degree $d$ is a positive
integer.   There is a theorem due to Tian and Yau which
states that as long as the index $I = \sum w_i - d$ is positive,
the cone $\Y$ cut out by $F=0$ is Calabi-Yau 
\cite{TianYau}.\footnote{
More precisely, the cone minus the apex $\bz=0$
is Calabi-Yau.}
We also insist that the only singularity of $\Y$ is at the
tip of the cone $r=0$.  This requirement on the 
singularities means the only solution to the system
of equations $\{\partial_i F = 0 : i = 0,1,2,3 \}$ is 
the point $\bz=0$.

In the previous section, we thought of $\Y$ as a fibration
over a K\"{a}hler-Einstein manifold $\bV$.  In terms of
$F$, we can think of $\bV$ as the corresponding variety
cut out by $F=0$ in weighted projective space
$\WP{w_0,w_1,w_2,w_3}$ rather than in affine
$\C{4}$.  Weighted projective space is defined in analogy to
ordinary projective space: instead of the uniform weighting,
the $\C{*}$-action on $\C{4}$ is weighted by a vector
of weights $(w_0, w_1, w_2, w_3)$.  The point
$\bz = 0$ is not included in $\WP{w_0,w_1,w_2,w_3}$.

In general, the space $\bV$ is not a smooth manifold but
rather has cyclic quotient singularities 
inherited from $\WP{w_0,w_1,w_2,w_3}$.  
In the coordinate patch $z_i \neq 0$,
weighted projective space looks like a quotient of 
$\C{3}$ by $\Z_{w_i}$.  
So there will be in general a quotient singularity
at the point $(0,0,z_i \neq 0,0)$.  
If the weights have common factors,
there may also be singular lines, planes, etc.  If
the variety $\bV$ 
intersects any of these singular regions, the variety
will also have quotient singularities.
To insure that the singularities of $\bV$ are of 
codimension 2 or less, one usually assumes that
the weighted projective space is well-formed:
\begin{eqnarray}
\label{wf1}
&\mathrm{gcd}(w_0,\ldots,\hat{w_i},\ldots,\hat{w_j},\ldots,w_n)
\mid d& \\
\label{wf2}
&\mathrm{gcd}(w_0,\ldots,\hat{w_i},\ldots,w_n) = 1&
\end{eqnarray}
The second condition is less stringent because any projective
space that does not satisfy condition (\ref{wf2}) is isomorphic to one
that does \cite{Dol,Fle}.

Although it is true that if $\bV$ is K\"{a}hler-Einstein,
then $\Y$ is Calabi-Yau, it does not appear to be true
in general that if $\Y$ is a Calabi-Yau cone that
there exists such a K\"{a}hler-Einstein $\bV$.  
Thousands of K\"{a}hler-Einstein $\bV$ have been
cataloged.  See for example \cite{JK1,JK2,BGN2}.
In this paper, we will be interested with a class of
$\bV$ where the corresponding $\Y$
are called generalized conifolds and for which the cone
$\Y$ is Calabi-Yau and smooth except at the tip, but
where it is not yet known whether $\bV$ admits a
K\"{a}hler-Einstein metric.  We will work under the
hypothesis that these $\bV$ are indeed K\"{a}hler-Einstein
and we will get sensible results for dibaryon masses.
  
\begin{figure}
\includegraphics[width=\textwidth]{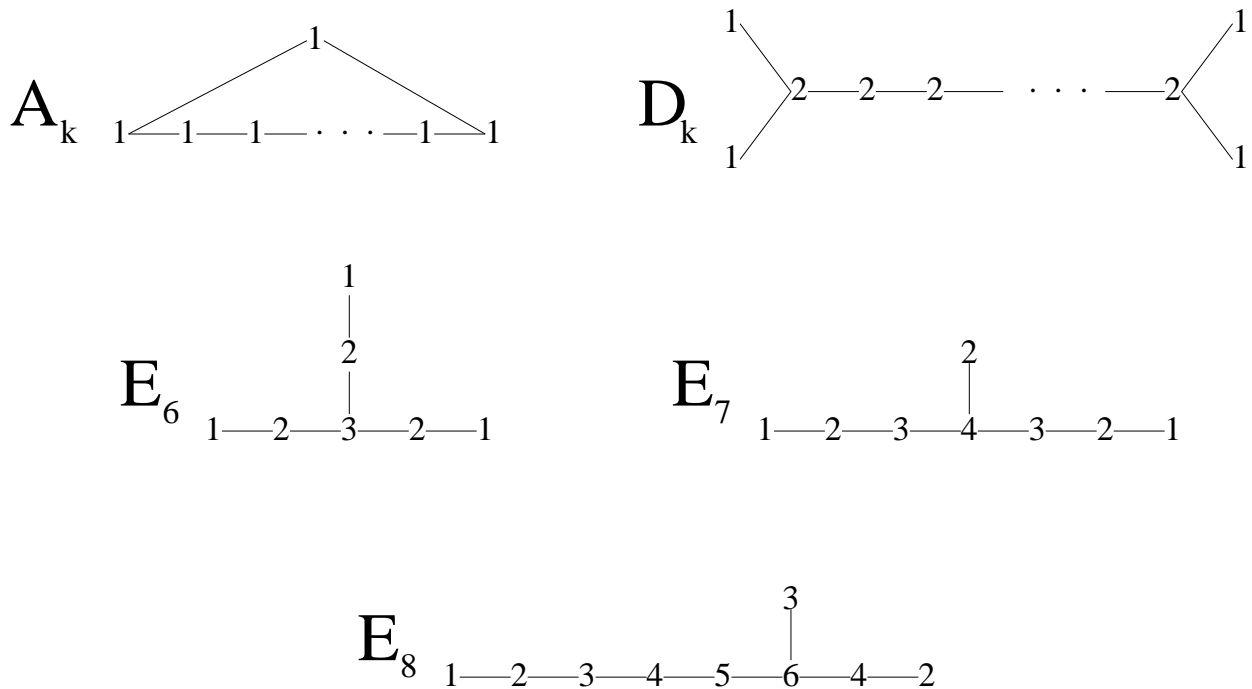}
\caption{The extended Dynkin diagrams of $ADE$ type, including the
         indices $n_i$ of each vertex.}
\label{fig1}
\end{figure}

\subsection{Gauge Theory Duals for Generalized Conifolds}

To introduce these generalized conifolds, 
let us begin by reviewing the gauge theory on the world volume
of a collection of 
$N$ D-branes placed at the orbifold singularity of $\C{2}/\Gamma$, where $\Gamma$ is a 
discrete subgroup of $SU(2)$ of $ADE$ type.\footnote{Much of this
discussion is drawn from \cite{GNS}.}  The field theory has ${\cal N}=2$ 
supersymmetry.  Its gauge group is the product
\[
G = \prod_{i=0}^{r}U(N_i)
\]
where $i$ runs through the set of vertices of the extended Dynkin diagram of the 
corresponding $ADE$ type (see figure \ref{fig1}) \cite{DM}.  
We have also introduced $N_i = N n_i$, where
$n_i$ is the index of the $i$th vertex of the Dynkin diagram.  Equivalently,
one may think of $i$ as running through the irreducible 
representations ${\bf r}_i$
of $\Gamma$, in which case $n_i$ can be thought of as the dimension
of  ${\bf r}_i$.

The field content of the gauge theory can be summarized 
conveniently with a 
quiver diagram which is in fact the corresponding extended Dynkin diagram.  
For each vertex in the Dynkin diagram, we have an ${\cal N}=2$
vector multiplet transforming under the adjoint of $U(N_i)$.  For each
line in the diagram, there is a bifundamental hypermultiplet $a_{ij}$ in
the representation $(N_i, \bar N_j)$.

To write a superpotential for this gauge theory, it is convenient to
decompose the fields into ${\cal N}=1$ multiplets.  Each $a_{ij}$ will
give rise to a pair of chiral multiplets, $(B_{ij}, B_{ji})$, where
$B_{ij}$ is a complex matrix transforming in the $(N_i, \bar N_j)$ 
representation.  Moreover, there is a chiral multiplet $\phi_i$ 
for each vector multiplet in the theory.

The superpotential is then
\be
W = \sum_{i} \Tr \mu_i \phi_i 
\label{suppot}
\ee
where $\mu_i$ is the ``complex moment map"
\be
{{\mu_i}^{\alpha_i}}_{\beta_i} = 
\sum_j s_{ij} {{B_{ij}}^{\alpha_i}}_{\gamma_j}
{{B_{ji}}^{\gamma_j}}_{\beta_i} \ .
\label{mmap}
\ee
Although the indices are confusing, essentially all we have done is construct
a cubic polynomial in the ${\cal N}=1$ superfields consistent with
${\cal N}=2$ SUSY and the gauge symmetry.  The factor $s_{ij}$ is the 
antisymmetric adjacency matrix for the Dynkin diagram: $s_{ij}=\pm 1$ when
$i$ and $j$ are adjacent nodes and zero otherwise.  The upper index $\alpha_i$
indicates a fundamental representation of $U(N_i)$, while a lower index
$\beta_i$ indicates an anti-fundamental representation of $U(N_i)$.
There is a relation among the $\mu_i$
\be
\sum_i \Tr \mu_i = 0 \ .
\label{consmu}
\ee 
which holds because the trace gives something
symmetric in $i$ and $j$ summed against $s_{ij}$ which is
antisymmetric.

This ${\cal N}=2$ gauge theory is superconformal and thus must have an
$R$ symmetry.
$W$ must have $R$ charge $2$.  Conveniently, the $B_{ij}$ and the $\phi_i$ have
$R$ charge $2/3$ and the superpotential, as noted above, is cubic. 

In the large $N$ limit in the case of D3-branes, we can invoke the AdS/CFT 
correspondence for this gauge theory \cite{LNV, KS}.  The correspondence
tells us that the gauge theory described above is dual to type IIB
supergravity (SUGRA) on an $AdS_5 \times \S^5/\Gamma$ 
background.  To see how the
orbifolding works, consider 
$\S^5 = \{(z_1, z_2, z_3)\in \C{3}: \sum_i |z_i|^2=1\}$.
The group $\Gamma$ acts only on $(z_1, z_2) \in \C{2}$.  As a result, 
there is an $\S^1$ of the $\S^5$ which is left invariant under $\Gamma$.

We can add a term to the superpotential
(\ref{suppot}) that will give masses $m_i$ to the $\phi_i$.  Such a term
will eliminate the $\phi_i$ from the theory
at energies below the mass scale set by the $m_i$ and 
break the supersymmetry from ${\cal N}=2$ to ${\cal N}=1$.  In particular,
we add the term
\[
W' = W - \frac{1}{2} \sum_i m_i \Tr \phi_i^2 \ .
\]  
%To see what happens at low energies, let us look at the equations of motion
%$dW'=0$.  By varying with respect to the matter fields that define $\mu_i$,
%we see that 
%\be
%\phi_i = \phi \Id_{N_i} 
%\label{phii}
%\ee
%where $\phi$ is the Lagrange multiplier used to ensure
%that the constraint (\ref{consmu}) is satisfied.
%Varying with respect to $\phi_i$ and employing (\ref{phii}), one gets
%\[
%\mu_i = m_i \phi \Id_{N_i} \ .
%\]
%{}From (\ref{consmu}), it follows that
%\[
%\sum_i n_i m_i = 0 \ .
%\]
Assuming that none of the $m_i=0$, we integrate out the $\phi_i$ from the action
to find an effective low energy superpotential:
\[
W_{eff} = \sum_i \frac{1}{2m_i} \Tr \mu_i^2 \ .
\]
Notice that $W_{eff}$ is quartic in the superfields $B_{ij}$ 
(see (\ref{mmap})).  
We would like the endpoint of the RG flow 
generated by adding
these mass terms to be an IR conformal fixed point.  
However, if the fields $B_{ij}$ are given
their naive $R$ charges of $2/3$, this quartic superpotential will 
explicitly break our $R$ symmetry.  By giving the $B_{ij}$ anomalous dimensions,
we find that after flowing to the IR, the $R$ charge of the $B_{ij}$ can
be adjusted to $1/2$, and the $R$ symmetry is preserved.  Thus,
the conformal dimension of the $B_{ij}$ is always 3/4.

\subsection{The Geometry of the Generalized Conifolds}

In the context of the AdS/CFT correspondence, the authors of \cite{GNS} generalized an
argument of \cite{KW} for the $A_1$ case, arguing that the IR endpoint of this RG flow is
dual to type IIB SUGRA in an $AdS_5 \times \bX_\Gamma$ background where the $\bX_\Gamma$
are the level surfaces of certain ``generalized conifolds''.\footnote{ Similar conclusions
were reached for the $A_k$ type generalized conifolds in \cite{EL}.}  The generalized
conifolds are three complex dimensional Calabi-Yau manifolds with a conical scaling
symmetry.  The conifolds can be described by a polynomial embedding relation $F_\Gamma=0$
in $\C{4}$.  To conform with the notation of \cite{GNS}, we use the coordinates
$(\phi,x,y,z)\in \C{4}$.  The polynomial $F_\Gamma$ is invariant under a $\C{*}$-action,
the real part of which is the conical scaling symmetry while the imaginary part
corresponds to an $R$ symmetry transformation in the dual gauge theory.  $F_\Gamma$
transforms under this $\C{*}$-action with weight $d$.  The coordinates $\phi$, $x$, $y$,
and $z$ transform with weights
\be
\renewcommand{\arraystretch}{1.2}
\begin{array}{cccccc}
\Gamma & [\phi] & [x] & [y] & d/2 = [z] & d \\
A_{2q} & 2 & 2 & 2q+1 & 2q+1 & 4q+2 \\
A_{2q+1} & 1 & 1 & q+1 & q+1 & 2q+2 \\
D_k & 1 & 2 & k-2 & k-1 & 2(k-1) \\
E_6 & 1 & 3 & 4 & 6 & 12 \\
E_7 & 1 & 4 & 6 & 9 & 18 \\
E_8 & 1 & 6 & 10 & 15 & 30 
\end{array}
\label{weights}
\ee
To make things more concrete, we give the polynomials 
\be
F_{A_k} = \prod_{i=0}^{k} (x - \xi_i \phi) + y^2 + z^2 \ ,
\label{ak}
\ee
\be
F_{D_k} = \prod_{i=0}^{k-2} (x - \xi_i \phi^2) + 
                           t_0 \phi^k y + xy^2 + z^2 \ ,
\ee
\be
F_{E_6} = y^3 + t_0 \phi^2 (x- a_1 \phi^3) (x-a_2 \phi^3) y +
\prod_{i=1}^4 (x - b_i \phi^3) + z^2 \ ,
\ee
\be
F_{E_7} = y^3 + (x - a_1 \phi^4)(x-a_2 \phi^4)(x-a_3\phi^4) y
+ t_0 \phi^2 \prod_{i=1}^4 (x - b_i \phi^4) + z^2 \ ,
\ee
\be
F_{E_8} = y^3 + t_0 \phi^2 (x - a_1 \phi^6)(x-a_2 \phi^6)(x-a_3\phi^6) y 
+ \prod_{i=1}^5 (x- b_i \phi^6) + z^2 \ ,
\ee
where $\xi_i$, $t_0$, $a_i$, and $b_i$ 
are free constants transforming with
weight zero.

 Our object is to calculate geometrically the quantity given in (\ref{result}),  
$$
\Delta = - 3N \frac{K_V\cdot C}{K_V \cdot K_V} \ .
$$
In fact we will see below that it suffices to minimize $\Delta$.  Thus it suffices to determine
\be 
-K_V\cdot K_V \label{k2}
\ee
and to minimise 
\be
-K_V\cdot C \label{object},
\ee
where $C$ ranges over all holomorphic curves in $V$.  We we will refer to any such curve on $V$ 
as a curve of minimal degree.  We note that in each case 
$V$ admits a double cover of a weighted projective space $W$ of dimension two,
$$
\pi\colon\map V.W.
$$
which corresponds to projection onto the $(\phi,x,y)$-coordinates from the point
$[0:0:0:1]$ of the corresponding weighted projective space.  The fact that $\pi$ is a
double cover corresponds to the fact that if we fix the value of $(\phi,x,y)$, then $z$
takes on two possible values, corresponding to the two choices of square root, positive
and negative.  

 Let $B$ be the branch locus of $\pi$.  Let $G_{\Gamma}$ be the polynomial in
$(\phi,x,y)$, obtained by setting $z=0$ in $F_{\Gamma}$.  Clearly the zero locus $\Sigma$
of $G_{\Gamma}$ is part of the branch locus of $\pi$.  However, in some of the cases when
we drop the coordinate $z$, the resulting weights are not well-formed.  In fact sometimes
two of the first three entries share a common factor of $2$.  The weights for $W$ are
obtained by canceling the common factor.  However it is easy to see, from the description
of weighted projective space as a quotient of $\mathbb{C}^4$, that in fact the coordinate
axis corresponding to the vanishing of the other coordinate is also a component of $B$.

 We let $e$ denote the degree of $B$.  Clearly the degree of $\Sigma$ is equal to the
degree of $G$, which is of course $d$.  It will be easy to calculate the degree of the
extra component, should it be present, and of course $e$ is nothing but the sum of these
degrees.

 Our aim is to reduce the calculation of the relevant intersection numbers on $V$ to a
calculation on $W$, which it will turn out is considerably easier.  To this end, we want
to apply the push-pull formula.  Given a cohomology class $\alpha$ on $V$ and a cohomology
class $\beta$ on $W$, push-pull reads as
$$
\pi_*(\alpha\cdot \pi^*\beta)=\pi_*\alpha\cdot \beta.
$$
 To apply push-pull then, we need to express $K_V$ as the pull-back of some class
from $W$.  In fact it is easy to do so, using the Riemann-Hurwitz formula, which 
in our case reads as
\be
K_V=\pi^*(K_W+1/2B).
\ee 

 Given this, it is easy to compute the number $K_V\cdot K_V$ on $W$,
\begin{align*} 
\pi_*(K_V\cdot K_V) &= \pi_*(\pi^*(K_W+1/2B)\cdot \pi^*(K_W+1/2B)\cdot [V])\\ 
                    &= \pi_*(\pi^*((K_W+1/2B)\cdot (K_W+1/2B))\cdot [V])\\ 
                    &= (K_W+1/2B)\cdot (K_W+1/2B)\cdot \pi_*[V]\\ 
                    &= (K_W+1/2B)\cdot (K_W+1/2B)\cdot 2[W]\\ 
                    &= 2(K_W+1/2B)\cdot (K_W+1/2B),\\ 
\end{align*} 
where we use the obvious geometric fact that $\pi_*[V]=2[W]$ (note also that the class of
$V$ is the identity in cohomology).  Thus 
\be 
K_V\cdot K_V=2(K_W+1/2B)\cdot (K_W+1/2B).
\ee

 Suppose that we are given a curve $C$ in $V$.  We want to do the same thing with $K_V\cdot
C$.  Let $C'$ be the image of $C$ inside $W$.  By definition the push-forward of $C$ is a
multiple of $C'$,
$$
\pi_*C=fC',
$$ 
where $f$ is the degree of the map $\pi|_C\colon\map C.C'.$.  As $\pi$ itself has
degree two, either $f=1$ or $2$, and we may distinguish the two cases by considering the
inverse image of $C'$.  $f=1$ if either the inverse image of $C'$ is the union of two
irreducible components (possibly connected), or $C'$ is a component of the branch locus
$B$.  Otherwise $f=2$, in which case $C$ is the inverse image of $C'$, but $C'$ is not a
component of the branch locus.  Now we can compute the intersection number $K_V\cdot C$ by push-pull
\begin{align*} 
\pi_*(K_V\cdot C) &= \pi_*(\pi^*(K_W+1/2B)\cdot C)\\ 
                  &= (K_W+1/2B)\cdot \pi_*C\\ 
                  &= f(K_W+1/2B)\cdot C'.\\ 
\end{align*} 

 Thus
\be
-K_V\cdot C=-f(K_W+1/2B)\cdot C'.
\ee

 Of course one tricky thing about this formula is that the value of $f$ depends on $C$.  
In this way, we reduce the problem of minimizing the intersection number $-K_V\cdot C$,
where $C$ ranges over all holomorphic curves in $V$, to minimizing the intersection number
\be 
-f(K_W+1/2B)\cdot C \label{easier} 
\ee 
where now $C$ ranges over all holomorphic curves in $W$.  In each case, it is not too hard to 
prove that given $C$, we may find a curve $D$ numerically equivalent to $kC$ such that the inverse
of $D$ represents $k\Delta$.  Thus it does indeed suffice to minimize (\ref{easier}).  
Consider the following table:
\be \renewcommand{\arraystretch}{1.2}
\begin{array}{ccccccccc}
\Gamma    & [\phi] & [x]          & [y]       & d    &  e    &  f   &\Delta/3N & \#\\
A_{2q}   &   1    &      1       & 2q+1      & 2q+1 &  4q+2 &  1   & 1/4      & 2(k+1)  \\
A_{2q+1} &   1    &      1       & q+1       & 2q+2 &  2q+2 &  1   & 1/4      & 2(k+1) \ .   \\
D_{2q}   &   1    &      1       & q-1       & 2q-1 &  2q   &  1   & 1/2      &   \\
D_{2q+1} &   1    &      2       & 2q-1      & 4q   &  4q   &  2   & 1/2      &   \\
E_6      &   1    &      3       &   4       & 12   &  12   &  2   & 1/2      &   \\
E_7      &   1    &      2       &   3       & 9    &  10   &  1   & 1/2      &   \\
E_8      &   1    &      3       &   5       & 15   &  16   &  1   & 1/2      &  
\end{array}
\label{new_weights}
\ee
 
 We next explain how we got the last four columns.  To proceed further, observe that any
weighted projective space is in fact an example of a toric variety.  Recall that a variety
is said to be a toric variety, if it contains a dense open subset isomorphic to a torus,
that is a copy of $(\mathbb{C}^*)^n$, where, moreover, the natural action of the torus
extends to an action on the whole of the toric variety.  Suppose that $X$ is a toric
variety and that $V$ is any subvariety.  We claim that the class $[V]$ of $V$ (either in
cohomology or better yet in the Chow ring) is an integral linear combination of classes of
invariant subvarieties (under the action of the torus), that is
$$
[V]=\sum _i a_i[Z_i] \label{general}
$$
where each $a_i$ is a non-negative integer and $Z_i$ ranges over the invariant
subvarieties.  Indeed if $V$ is not already invariant, then we may find a one dimensional
torus $\mathbb{C}^*$ inside the big torus, which moves $V$ inside $X$.  Taking the limit,
we obtain a degeneration of $V$ to a collection of subvarieties that are invariant under a
subgroup of the torus of larger dimension.  Continuing in this way, we finally degenerate
$V$ to a sum of cycles, all of which are invariant under the action of the whole torus.
 
 In our case $C$ is a curve and $W$ is a toric surface of Picard number one.  On a toric
surface, the only invariant subvarieties are the surface itself, a finite union of
invariant curves and their intersection points.  As $C$ is a curve, we only need worry
about the invariant curves, and as $W$ has Picard number one, there are only three
invariant curves, $B_0$, $B_1$ and $B_2$ say, which form a triangle, where each invariant
curve is given as the vanishing of one of the coordinates, $\phi$, $x$ or $y$.  Thus
(\ref{general}) reduces to \be [C]=\sum a_i[B_i], \ee where each $a_i$ is a non-negative
integer.  Thus to minimize $-(K_W+1/2B)\cdot C$, it suffices to find the minimum of
$-f(K_W+1/2B)\cdot B_i$, for $i=0$, $1$ and $2$.  The condition that the Picard number is
one means that the cohomology classes of any two curves are proportional.  In particular
$$
B_i=\lambda_{i,j}B_j,
$$ 
where summation notation has been suppressed. Clearly we want to choose $j$, so that for
all $i$, $\lambda_{i,j}\geq 1$.  To determine the constants, it suffices to compare the
intersection numbers $B_i\cdot B_j$.  Suppose the weights are $(w_0,w_1,w_2)$, so that
$W=\WP{w_0,w_1,w_2}$.  By symmetry it suffices to consider the case $B_0\cdot B_1$.  Now
$W$ has three invariant points, the vertices of the triangle, that is the points
$p_2=B_0\cap B_1$, $p_1=B_0\cap B_2$ and $p_0=B_1\cap B_2$.  Thus to calculate the
intersection number $B_0\cdot B_1$ it suffices to compute the local intersection number at
the point $p_2$.  Locally we have
$$
f\colon\map \mathbb{C}^2.\mathbb{C}^2/\mathbb{Z}_r.,
$$ 
where $r=w_2$ is the index of the singular point.  Let $C_0$ and
$C_1\subset\mathbb{C}^2$ be the axes upstairs.  Then the local intersection number
$C_0\cdot C_1=1$ and of course $C_1=f^*B_1$.  Hence, by push-pull
\begin{align*} 
1 &= f_*(C_0\cdot C_1) \\
  &= f_*(C_0\cdot f^*B_1) \\
  &= f_*(C_0)\cdot B_1) \\
  &= rB_0\cdot B_1, \\ 
\end{align*} 
where we use the fact that $f_*C_0=rB_0$, as $\map C_0.B_0.$ is an $r$-fold cover.  

 Thus 
\be
B_0\cdot B_1=1/w_2 \qquad B_0\cdot B_2=1/w_1 \qquad \text{and} \qquad B_1\cdot B_2=1/w_0.
\ee
 
 Putting all of this together we get 
\be 
\lambda_{i,j}=\frac {w_i}{w_j}.  
\ee 

 In our case, $w_0=1$, and the weights are increasing.  Thus every curve $C$ in $W$ is
numerically equivalent to an integral multiple of $B_0$.  Thus it is natural to express
all curves as multiples of $B_0$.  

 As $f=1$ or $2$, it follows that a curve $C$ numerically equivalent to $B_0$ minimizes
(\ref{easier}), so that we are reduced to finding the minimum of 
\be 
f(K_W+1/2B)\cdot C \label{trivial}, 
\ee 
where $C$ is a curve numerically equivalent to $B_0$.  For $D_{2q}$, $E_7$ and $E_8$,
$B_0$ is a component of $B$.  Thus $f=1$, by definition.  For $A_k$, $B_0$ and $B_1$ are
numerically equivalent.  In fact $W$ is a cone over a rational normal curve, and $B_0$ and
$B_1$ correspond to lines of the ruling.  Let $L$ be any line of this ruling and let $M$
be the inverse image of $L$.  $L$ is determined by setting a linear combination of $\phi$
and $x$ to zero (for example $B_0$ corresponds to $\phi=0$ and $B_1$ to $x=0$), so that we
fix the ratio between $\phi$ and $x$.  Having fixed this ratio, then $M$ is a conic in a
weighted projective plane (in fact the cone over $L$ and the point of projection), given
on an affine piece by an equation of the form
$$
z^2+y^2=l,
$$ 
where $l$ is a constant.  For us the only important thing is whether $l$ is zero or not.
If $l$ is zero, then $z^2+y^2$ factors as the product of two linear polynomials, and $M$
consists of two lines.  Otherwise $M$ is a smooth conic.  Of course $l$ is zero iff
$x=\xi_i\phi$, so that this happens $k+1$ times.  In this case $f=1$ as well.  Moreover
there are two $2(k+1)$ curves of minimal degree, two for each $i$,  
giving the last
column of the first and second rows 
of (\ref{new_weights}).  In the other two cases, $D_{2q+1}$ and
$E_6$, $B_0$ is the only (effective) curve in its numerical equivalence class, and the
inverse image of $L$ is easily seen to be irreducible and so $f=2$ in these two cases.
These values for $f$ give the second from last column of (\ref{new_weights}).

 We turn to the problem of expressing the classes of $K_W$ and $B$ in terms of $B_0$.  It
is a basic property of any toric variety $X$ that $-K_X$ is equivalent to a sum of all the
invariant divisors.  In our case, it is particularly easy to see that this is true; the
invariant curves $D=B_0+B_1+B_2$ form a triangle, and each is a copy of $\pr 1.$.  Taken
together they are then a curve of genus one, and so $K_D=0$.  But by adjunction 
$$
(K_W+D)|_D=K_D=0.
$$
As we are on a surface of Picard number one, the fact that the restriction is zero implies that 
$K_W+D=0$.  Thus
\begin{align*} 
-K_W &= B_0+B_1+B_2\\ 
     &= \left (1+\frac {w_1}{w_0}+\frac {w_2}{w_0}\right )B_0,\\ 
\end{align*} 
so that
\be
-K_W=\frac 1{w_0}\left (w_0+w_1+w_2\right )B_0.
\ee
 To compute the class of $B$ note that in all cases, $B_0$ has weight one.  It follows that 
\be
B=eB_0,
\ee
where $e$ is the degree of $B$.  In all cases when $e\neq d$, excepting the case $A_{2q}$, the 
extra component of the branch locus is $B_0$ itself.  In these cases $e=d+1$.  Otherwise in 
the case of $A_{2q}$, the extra component is $B_2$ and $B_2=(2q+1)B_0$, so that $e=d+(2q+1)=4q+2$.  

 Finally we note one further advantage of working on $W$, a surface of Picard number one.  To compute 
the ratio
$$
\frac {-f(K_W+1/2B)\cdot C}{2(K_W+1/2B)\cdot (K_W+1/2B)}
$$
we note that when we express everything in terms of $B_0$, there will be a lot of canceling.  In fact 
it is easy to see that this ratio reduces to 
$$
\frac f{2\lambda}
$$
where $-(K_W+1/2B)=\lambda C$.  In our case, we take $C=B_0$.  Now
$$
-(K_W+1/2B)=(1+w_1+w_2-e/2)B_0,
$$
where we use the fact that $w_0=1$, so that
$$
2\lambda=2(1+w_1+w_2-e/2)=2+2w_2+2w_2-e.
$$
Putting all of this together, we obtain that $\Delta/3N$ will always be an integer
multiple of the penultimate column of (\ref{new_weights}).

\subsection{Dibaryons from Gauge Theory}

Paths in the simply laced
Dynkin diagrams of figure \ref{fig1} correspond
to dibaryonic operators.  
In particular, pick two nodes of a simply
laced Dynkin diagram corresponding to gauge groups 
$SU(j N)$ and $SU(k N)$.  To construct the dibaryon, we
first need to construct a smaller object transforming
in the bifundamental of $SU(jN) \times SU(kN)$.  Choose
a path along the Dynkin diagram that joins these two nodes.
The path can double back on itself.  
For each link joining
two nodes in the path, write down the corresponding 
bifundamental matter field.  Which bifundamental we choose
depends on which direction we move between the nodes.
At this point, we will have some number $s$ of 
bifundamental matter fields where we can trace over
every index save the fundamental of $SU(jN)$ and
the antifundamental of $SU(kN)$.  Call this object
of conformal dimension $3s/4$ ${\mathcal O}^{\alpha}_{\beta}$.  

Now if $j=k$, then we can antisymmetrize over $jN$ copies
of ${\mathcal O}$ and the conformal dimension of the dibaryon
will simply be 
\be
\Delta = \frac{3sN}{4} j \ .
\ee

If $j\neq  k$, we have to be more careful, as we saw in 
the case of the fourth del Pezzo.   
We need to choose some more paths in the Dynkin
diagram, corresponding to some collection of
bifundamental fields ${\mathcal O}_i$.  
If we choose the paths correctly, we can 
antisymmetrize over the collection
to form a (non-zero) gauge invariant
dibaryon.    
For $j\neq k$, one bad idea is to antisymmetrize over
$\lcm(j,k) N$ copies of the original field
${\mathcal O}$.  Such an antisymmetric sum
is zero.  However, if we can construct
a collection of fields with the same
path length $s$ such that all transform
under $SU(jN) \times SU(kN)$ and such
that the antisymmetrization over each
$SU(jN)$ and $SU(kN)$ is nonzero, then
we find the rather remarkable formula
\be
\Delta = \frac{3sN}{4} \lcm(j,k) \ .
\ee

Staring at the Dynkin diagrams, it is straightforward to
compute the different possible values of the 
conformal dimension.  For the $A_k$ types, the dimension
is an integer multiple of $3N/4$.  For the
$D_k$ and $E_k$ types, the dimension is an integer multiple
of $3N/2$.  
These numbers are exactly as predicted by the geometric
calculation presented above.

We may also count the number of dibaryonic
operators of smallest conformal dimension.  
For the $A_k$ type quivers,
there is a smallest dibaryon for each elementary 
bifundamental
matter field $B_{ij}$, 
or equivalently for each path of length $s=1$.  Thus,
there are $2(k+1)$ smallest dibaryons for the
$A_k$ quiver,
in precise agreement with the 
geometric calculation presented above in the
last column of table (\ref{new_weights}).

\section{Remarks}

This paper marks a step forward in the authors' 
understanding of 
the relation between dibaryons in superconformal gauge
theories and holomorphic curves in K\"{a}hler-Einstein
surfaces all in the context of the AdS/CFT correspondence.
The formula (\ref{result}) is a
powerful way of calculating geometrically the conformal dimension
of time independent 
dibaryons in a wide variety of contexts without
having explicit knowledge of the metric.  
We applied
this formula to some well studied examples of 
AdS/CFT correspondence, namely the del Pezzos and
the generalized conifolds, where previously lack of a metric
had hampered progress.  In all cases, we found good agreement
with previously established gauge theory results.

Having remarked on the progress, there are a large number of
questions which still need to be addressed.  For example, 
how does the number of holomorphic curves $C$ with a given value of
$-K_\bV \cdot C$ compare to the number of dibaryons of a given
conformal dimension.  For the del Pezzos
and the $A_k$ type generalized conifolds, 
we were able to count the number of smallest dibaryons and 
compare successfully with the number of
lowest degree
holomorphic curves.  However, for the del Pezzo gauge theories,
there seem to be too many larger dibaryons.  
The superpotential provides some classical relations 
between the bifundamental matter fields that partially reduce this number.
Beasley and Plesser
demonstrated the existence of additional ``quantum'' relations between
bifundamental matter fields that reduced this naive number
even further for the third del Pezzo \cite{Beasley1}.  
Taking into account both classical and ``quantum'' relations, Beasley and
Plesser were able to match the number of gauge theory dibaryons with the number
of holomorphic curves.
It would be interesting
to reproduce their calculation for the other del Pezzos.  It would
also be interesting to count dibaryons 
more generally for the generalized conifolds.

A bizarre development is the agreement between gauge theory and geometry
for the conformal dimension
of dibaryons in the first and second del Pezzo.  The first and second del
Pezzo are known not to admit a K\"ahler-Einstein metric, and it is a little difficult to imagine how (\ref{result}) remains meaningful in this case.
Along the same lines, no one has yet demonstrated that the 
$\bV$ of the generalized
conifolds admit a K\"ahler-Einstein metric.  It would be interesting to know
whether these $\bV$ are K\"ahler-Einstein or whether they fall into the 
same category as the first and second del Pezzos.

Another interesting question is 
how Seiberg duality affects, or rather fails
to affect, the spectrum of dibaryons.  
Seiberg dual theories can have 
very different quivers and 
superpotentials.  Somehow, one continues to find
the same set of conformal dimensions for the dibaryons.

Finally, given the geometry $AdS_5 \times \bX$,
is it possible that the dibaryon spectrum can be used to
construct the gauge theory dual?  
In all the examples considered here, 
except the seventh and eighth del Pezzo, the gauge
theory duals had been constructed using independent considerations.  
In general, constructing these gauge theory duals is nontrivial,
and dibaryons may prove to be a useful additional tool. 

We hope to return to some of these questions in the future.

Note: While this letter was being prepared, we learned of
\cite{IW2} which overlaps this work to some extent.

\section*{Acknowledgments}
C.~H. would like to single out Igor Dolgachev for 
special thanks.  C.~H. would also like to thank
the MCTP in Ann Arbor, where this project was resurrected, for
hospitality.  
We would like to 
thank C.~Beasley and A.~Bergman for collaboration
in the early stages of this project.  We would also like
to thank A.~Iqbal, A.~Mikhailov, and J.~Walcher for
useful conversations.  This research was supported in 
part by the National Science Foundation under Grant No.
PHY99-07949.

\end{document}